\documentclass[12pt]{iopart}
\usepackage{amssymb}
\usepackage[usenames,dvipsnames]{color}
\begin{document}
\title{An anisotropic standing wave braneworld and associated Sturm--Liouville problem}

\author{Merab Gogberashvili$^1$, Alfredo Herrera--Aguilar$^{2,3}$ and
Dagoberto Malag\'on--Morej\'on$^2$}

\address{$^1$Andronikashvili Institute of Physics, 6 Tamarashvili St., Tbilisi 0177, Georgia \&\\
~Javakhishvili State University, 3 Chavchavadze Ave., Tbilisi 0128, Georgia}

\address{$^{2}$Instituto de F\'{\i}sica y Matem\'{a}ticas, Universidad Michoacana de
San Nicol\'as de Hidalgo, Edificio C--3, Ciudad Universitaria, C.P. 58040, Morelia, Michoac\'{a}n,
M\'{e}xico}

\address{$^{3}$Centro de Estudios en F\'{\i}sica y Matem\'{a}ticas B\'{a}sicas y Aplicadas, Universidad
Aut\'{o}noma de Chiapas, Calle 4a Oriente Norte 1428, Tuxtla Guti\'{e}rrez, Chiapas, M\'{e}xico}

\eads{gogber@gmail.com, alfredo.herrera.aguilar@gmail.com, malagon@ifm.umich.mx}

\date{\today}

\begin{abstract}
We present a consistent derivation of the recently proposed 5D anisotropic standing wave
braneworld generated by gravity coupled to a phantom-like scalar field. We explicitly solve the
corresponding junction conditions, a fact that enables us to give a physical interpretation to the
anisotropic energy-momentum tensor components on the brane. So matter on the brane represents an
oscillating fluid which emits anisotropic waves into the bulk. We also analyze the Sturm-Liouville
problem associated to the correct localization condition of the transverse to the brane metric and
scalar fields. It is shown that this condition restricts the physically meaningful space of
solutions for the localization of the fluctuations of the model.
\end{abstract}

\maketitle


\section{Introduction}

Since the early proposals of braneworld models involving large extra dimensions and 4D
delta-function sources with both positive and negative tensions \cite{Hi,brane} there has been a
lot of activity in this area with the aim of solving several open questions in modern physics (see
\cite{reviews,maartenskoyama} for reviews). Most of these models were realized as time independent
field configurations. However, quite soon, mostly within cosmological approaches, there have
appeared several braneworlds that assumed time-dependent metrics and fields in an attempt to
address several open problems in astrophysics and cosmology \cite{S}-\cite{vartensions}. In
general, these braneworld models have a series of remarkable features: they develop a novel
geometrical mechanism of dimensional reduction based on a curved extra dimension, provide a
realization of the AdS/CFT correspondence to lowest order, take into account the self-gravity on
the brane through its tension, can trap various matter fields (except for gauge bosons) inside the
brane, recast the hierarchy problem into a higher dimensional viewpoint, and also lead to
cosmological backgrounds with consistent dynamics that recover general relativity results under
suitable restrictions on their parameters. All these facts provide a complex but interesting
interplay between gravity, particle physics and geometry \cite{maartenskoyama}.

However, there are still several open relevant questions which deserve attention: construction of
the simplest realistic solution for an astrophysical black hole on the brane and study its
physical properties like staticity and Hawking radiation, development of realistic approximation
schemes and numerical codes to study cosmological perturbations on all scales, computation of the
CMB anisotropies and large scale structure to confront their predictions with high-precision
observations \cite{maartenskoyama}.

One of the main drawbacks of the delta--function braneworld models from the gravitational
viewpoint is related to their singularities at the positions of the branes, a fact that gets worse
in models with more than six dimensions due to the stronger self-gravity of the brane. In an
attempt to heal this problem, several authors smooth out the brane tensions \cite{Csaki} or
introduced single \cite{gremm} or several scalars \cite{intscalars}, and phantom \cite{phantom} or
tachyon \cite{tachyon} fields as sources (for a recent review see \cite{thickreview}), and even
considered modified gravity braneworlds \cite{LiufR}. However, scalar field thick brane
configurations with 4D Poincar\'e symmetry also develop naked singularities at the boundaries of
the bulk manifold if a mass gap is required to be present in the graviton spectrum of Kaluza-Klein
fluctuations (see \cite{gremm,hammmln}, for instance). An alternative model without such a
drawback is provided by a recently proposed de Sitter braneworld model purely generated by 4D and
5D cosmological constants \cite{cuco}.

In this sense, it is important to consider new generalizations that attempt to make more realistic
the original braneworld models, or explore other aspects of higher-dimensional gravity which are
not probed or approached by these simple models.

In this paper we shall consider the braneworld generated by 5D anisotropic standing gravitational
waves coupled to a phantom-like scalar field in the bulk as it was recently proposed in
\cite{Gog1}. It turns out that standing wave configurations can provide a natural alternative
mechanism for localizing 4D gravity as well as for trapping matter fields (for scalar and fermion
fields see \cite{gmmscalar} and \cite{gmmfermion}, respectively), including gauge bosons
\cite{gmmboson}, which usually are not localized on thin braneworlds. A peculiarity of this
anisotropic braneworld scenario is that it possesses non-stationary metric coefficients in the 4D
part of the line element. We further impose $Z_2$-symmetry along the extra dimension, a fact that
gives rise to the need of introducing as well an anisotropic brane energy-momentum tensor for the
self-consistency of the model. When the corresponding junction conditions are solved, this fact
allows us to give a physical interpretation to the components of the anisotropic energy-momentum
tensor of the brane. Then the localization of transverse scalar and metric fields are treated
through the associated Sturm-Liouville method that restricts the physically meaningful parameter
space of the solution. Some final remarks are given at the end of the paper.


\section{The model}

In this section we briefly recall the 5D standing wave braneworld model proposed in \cite{Gog1},
which is generated by gravity coupled to a non self-interacting scalar phantom-like field
\cite{phantom,thickreview} which depends on time and propagates in the bulk, and is given by the
action:
\begin{equation} \label{action}
S_b = \int d^5x \sqrt{g}\left[\frac{1}{16 \pi G_5} \left( R - 2\Lambda_{5}\right)+
\frac{1}{2}\left(\nabla \phi\right)^2\right]~,
\end{equation}
where $G_5$ and $\Lambda _5$ are 5D Newton and cosmological constants, respectively. It is worth
noticing that in order to avoid the well-known problems of stability which occur with ghost
fields, the phantom-like scalar field does not couple to ordinary matter in the model.

The Einstein equations for the action (\ref{action}) read:
\begin{equation} \label{field-eqns}
R_{\mu \nu} - \frac{1}{2} g_{\mu \nu} R =  8\pi G_5 T_{\mu \nu} - \Lambda _5 g_{\mu \nu}~,
\end{equation}
where Greek indices run from $0$ to $5$, labeling the 5D space-time, and $T_{\mu \nu}$ is the
energy-momentum tensor for the scalar field
\begin{eqnarray} \label{EMT}
T_{\mu \nu} = -\partial_{\mu}\phi\partial_{\nu}\phi + \frac{1}{2}g_{\mu\nu}\partial_{\rho}\phi
\partial^{\rho}\phi~.
\end{eqnarray}

Following \cite{Gog1}, we use the anisotropic metric {\it ansatz}:
\begin{eqnarray} \label{metricA}
ds^2 = e^{2A(r)}\left( dt^2 - e^{u(t,r)}dx^2 - e^{u(t,r)}dy^2 - e^{-2u(t,r)}dz^2 \right) - dr^2~,
\end{eqnarray}
where Latin indices stand for the 4D space-time coordinates ($x^0=t$, $x^i$ ,with $i=1,2,3,$ are
the spatial coordinates $x,y,z$, respectively) and the extra dimension is denoted as $x^5=r$. This
metric generalizes straightforwardly the thin brane metric {\it ansatz} \cite{brane}, where $A(r)
\sim |r|$, which is recovered in the limit when $u(t,r)$ vanishes. The warp factor here $A(r)$ is
an arbitrary function of the extra coordinate $r$ and, in principle, may model a thick brane
configuration in the spirit of \cite{gremm}.

This braneworld constitutes a generalization of the thin brane model with the peculiarity that now
the brane possesses anisotropic oscillations on it, which send a wave into the bulk (as in
\cite{gms}), i.e. the brane is warped along the spatial coordinates $x, y, z$ through the factor
$e^{u(t,r)}$, depending on time $t$ and the extra coordinate $r$. Several anisotropic braneworld
models have been previously considered in the literature \cite{kasnerads}-\cite{GK} when
addressing relevant cosmological issues like anisotropy dissipation during inflation
\cite{anisotbwinfl}, braneworld isotropization with the aid of magnetic fields
\cite{bwisotropization} and localization of test particles \cite{GK}. Moreover, as a general
feature it has been established that anisotropic metrics on the brane prevent the bulk from being
static \cite{bcosmoanisotbulk,bwisotropization}.

The phantom-like scalar field $\phi (t,r)$ obeys the Klein-Gordon equation on the background
space-time given by (\ref{metricA}),
\begin{eqnarray} \label{phi}
\frac{1}{\sqrt{-g}}~\partial_\mu (\sqrt{-g} g^{\mu\nu}\partial_\nu \phi) =
\Box\phi = e^{-2A(r)}\ddot \phi - \phi'' - 4 A' \phi' = 0 ~,
\end{eqnarray}
where overdots mean time derivatives and primes stand for derivatives with respect to the extra
coordinate $r$.

We further write the Einstein equations in the form:
\begin{equation} \label{field-eqns1}
R_{\mu \nu} = - \partial_\mu \sigma\partial_\nu \sigma + \frac{2}{3} g_{\mu \nu} \Lambda _5~,
\end{equation}
where the gravitational constant has been absorbed in the definition of the scalar field:
\begin{equation}
\label{sigma} \sigma = \sqrt{8 \pi G_5} ~ \phi ~,
\end{equation}
and we have used the reduced form of the energy-momentum tensor of the phantom-like scalar field
\begin{equation} \label{Tmnbar}
\overline{T}_{\mu\nu} \equiv T_{\mu\nu} - \frac{1}{3}g_{\mu\nu}T
=-\nabla_{\mu}\sigma\nabla_{\nu}\sigma~,
\end{equation}
and $T\equiv g^{\mu\nu}T_{\mu\nu}$.

It turns out that from the $05$-component of the Einstein equations (\ref{field-eqns1}), it
follows that the fields $\sigma$ and $u$ are related by \cite{Gog1},
\begin{equation} \label{sigma=u}
\sigma (t,r) = \sqrt{\frac{3}{2}}~u(t,r)~,
\end{equation}
up to an additive constant that can be absorbed into a redefinition of the 4D spatial coordinates
in (\ref{metricA}).

On the other hand, by combining the $22$- and $33$-components of the Einstein equations
(\ref{field-eqns1}), and comparing the result to the corresponding $55$-component, one gets a
quite strong restriction on the function $A(r)$:
\begin{equation} \label{A''=0}
A'' = 0~,
\end{equation}
whose solution is linear in $r$, but corresponds to divergent warp factors. Another interesting
fact that follows from the same analysis is that this condition also implies a sort of fine
tuning, where the 5D cosmological constant is related to $A(r)$ as follows:
\begin{eqnarray} \label{L=a2}
\Lambda_5 = 6 {A'}^2~.
\end{eqnarray}
It is worth noticing that the relation (\ref{sigma=u}) remains valid even if we include a
self-interaction potential for the scalar field.

In order to get an asymptotically vanishing warp factor and to study a more interesting braneworld
with the above ($A'' = 0$) behavior on the bulk, we modify the initial model by imposing
$Z_2$-symmetry along the extra dimension and by introducing a thin brane at $r=0$, which, for
consistency with (\ref{metricA}), must be supported by an anisotropic energy-momentum tensor. This
implies that the warp factor of the metric (\ref{metricA}) adopts the known form, similar to
\cite{brane}:
\begin{eqnarray}\label{metric1}
ds^2 = e^{2a|r|}\left( dt^2 - e^{u}dx^2 - e^{u}dy^2 - e^{-2u}dz^2\right) - dr^2~,
\end{eqnarray}
where $a$ is an arbitrary constant. The non-zero components of the Ricci tensor for this metric
read:
\begin{eqnarray} \label{ricci}
R_{tt} &=& e^{2a|r|} \left[-\frac {3}{2} e^{-2a|r|} \dot u ^2 + 4a^2 + 2a\delta(r)\right]~, \nonumber \\
R_{xx}&=&R_{yy}=e^{2a|r|+u}\left[ \frac {1}{2} e^{-2a|r|}\ddot u -
4a^2 - 2a \epsilon(r) u' - 2a\delta(r)-\frac {1}{2} u'' \right]~, \nonumber \\
R_{zz} &=& e^{2a|r|-2u}\left[ - e^{-2a|r|}\ddot u - 4a^2 + 4a\epsilon(r)u' -2a\delta(r) + u''\right]~, \\
R_{rr} &=& -\frac 32 u'^2 - 4a^2 - 8a\delta(r)~, \nonumber\\
R_{rt} &=& -\frac 32 \dot uu' ~, \nonumber
\end{eqnarray}
where $\epsilon(r)$ is the sign function. The presence of terms proportional to $\delta(r)$ in the
Ricci tensor, together with the anisotropy of the metric, forces us to define an anisotropic
energy-momentum tensor on the brane $\tau^{\mu}_{\nu}$ with different stresses along different
directions. The simplest way to accomplish this is by proposing:
\begin{eqnarray} \label{tensormixto}
\tau^{\mu}_{\nu}=\delta{(r)}~\mbox{\rm diag}[\lambda_0,\lambda_1,\lambda_2, \lambda_3,0]~, ~~~~~
\lambda_1 = \lambda_2~.
\end{eqnarray}
The above energy--momentum tensor can be interpreted as certain 'anisotropic effective fluid'
which constitutes a mixture of a vacuum fluid (which is characterized just by a brane tension) and
an anisotropic matter fluid (see \cite{maartenskoyama} and references therein for a complete
decomposition of the energy-momentum tensor induced on the brane).

The parameter $\lambda_0$ is the energy density of the 'effective fluid' and $\lambda_m$
($m=1,2,3$) are the stresses along the $m$ directions (see below). In general these quantities
depend only on time. A similar energy-momentum tensor was implemented in the context of E\"otv\"os
braneworlds with variable (time-dependent) tensions in analogy with fluid membranes in
\cite{vartensions}.

By taking into account the above considerations, the field equations (\ref{field-eqns1}) change as
follows:
\begin{equation} \label{field-eqnsDELTA}
R_{\mu\nu} = - \partial_{\mu} \sigma\partial_{\nu} \sigma +
\frac{2}{3} g_{\mu\nu} \Lambda _5 + 8 \pi G_5 \bar{\tau}_{\mu\nu}~,
\end{equation}
where the reduced energy-momentum tensor,
\begin{equation}
\bar{\tau}_{\mu\nu} = \tau_{\mu\nu} - \frac{1}{3}g_{\mu \nu}\tau~,
\end{equation}
corresponds to the matter content on the brane and takes the form:
\begin{eqnarray}\label{reduced}
\bar{\tau}_{\mu\nu}&=&\frac{1}{3}\delta{(r)}~\mbox{\rm diag}
\left[(2\lambda_0-2\lambda_1-\lambda_3 ),
(\lambda_0-\lambda_1+\lambda_3)e^{u},\right.\\
&\,&\left.(\lambda_0-\lambda_1+\lambda_3)e^{u},
(\lambda_0+2\lambda_1-2\lambda_3)e^{-2u},
(\lambda_0+2\lambda_1+\lambda_3)\right]. \nonumber
\end{eqnarray}
By making use of (\ref{metric1}), (\ref{field-eqnsDELTA}) and (\ref{reduced}) one can show that
the relation (\ref{sigma=u}) between the scalar field and the metric function is satisfied as
well. The relation (\ref{L=a2}) computed with the aid of the metric (\ref{metric1}) adopts the
following form:
\begin{equation}
\label{L=a} \Lambda_ 5=6 a^2~.
\end{equation}
Therefore, the system of Einstein and field equations reduces to a single ordinary differential
equation,
\begin{eqnarray}\label{eqntou}
\label{field-eqnbulk} e^{-2a|r|}~\ddot u - u'' - 4a\epsilon(r)u' = 0~.
\end{eqnarray}

In addition to the latter, the existence of the thin brane in our setup gives rise to a
relationship between the jump of the first derivative of the function $u$ at $r=0$ (denoted by
$[u']$) the stresses $\lambda_m$ and the parameter $a$. A way to derive these relationships
consists in integrating (\ref{field-eqnsDELTA}) with respect to $r$ in the range $[-\varepsilon,
\varepsilon]$ and then making $\varepsilon$ tend to zero, yielding the following junction
conditions for the stresses $\lambda_m$:
\begin{eqnarray}
&\, a=\frac{4\pi G_5}{3}\left( 2\lambda_0 -2\lambda_1-\lambda_3\right)~, \nonumber \\
&[u']= \frac{16\pi G_5}{3}\left(-\lambda_0+\lambda_1-\lambda_3 \right)-4a~, \nonumber \\
&[u']= \frac{8\pi G_5}{3}\left(\ \lambda_0+2\lambda_1-2\lambda_3\right)+ 2a~, \label{match} \\
&\, a=-\frac{\pi G_5}{3}\left(
\lambda_0+2\lambda_1+\lambda_3\right)~. \nonumber
\end{eqnarray}
Although the above system seems to be overdetermined since we have four equations for just three
variables $\lambda_0$, $\lambda_1$ and $\lambda_3$, it is possible to show that only three
equations are independent.

We then construct a standing wave solution to (\ref{eqntou}) by implementing the {\it ansatz}:
\begin{equation} \label{separation}
u(t,r) = C  \sin (\omega t) f(r)~,
\end{equation}
where $C$ and $\omega$ are some constants. In \cite{Gog1} it was argued that a standing wave
configuration in the bulk would in principle provide an alternative mechanism for localizing
gravity and binding matter fields (through an anisotropic force which drives the matter
fields/particles toward the nodes), including light (see \cite{gmmscalar}-\cite{gmmboson} for
recent results).

The equation for the radial function $f(r)$ from (\ref{separation}) reads,
\begin{equation} \label{f}
f'' + 4 a\epsilon(r) f' + \omega^2 e^{-2 a |r|}f = 0 ~.
\end{equation}
The general solution of this equation adopts the following form:
\begin{eqnarray} \label{fsol} f(r) = e^{-2a|r|} \left[{\cal A}~J_2\left(
\frac{\omega}{|a|} e^{-a|r|} \right) + {\cal B}~Y_2\left( \frac{\omega}{|a|} e^{-a|r|}
\right)\right],
\end{eqnarray}
where ${\cal A},{\cal B}$ are arbitrary constants and $J_2$ and $Y_2$ are $2^{nd}$ order Bessel
functions of the first and second kind, respectively. It is worth noticing that although the $Y_2$
Bessel function is singular as $r \longrightarrow \infty$ for $a > 0$, the whole function
(\ref{fsol}) vanishes in this limit since its behaviour is dominated by the $e^{-2a|r|}$ factor.

As pointed out in \cite{Gog1}, along with the solution (\ref{fsol}), we impose the ghost-like
field to be unobservable on the position of the thin brane, since it oscillates together with the
gravitational field (see (\ref{sigma=u})). A way to accomplish this consists in setting a boundary
condition that nullifies the ghost-like field at the position of the brane $r=0$, a condition that
quantizes the oscillation frequency, $\omega$, of the standing wave:
\begin{equation} \label{quantize}
\frac{\omega}{|a|} = X_{n}~,
\end{equation}
where $X_{n}$ is the $n^{th}$ zero of the $2^{nd}$ order Bessel function $J_2$ or $Y_2$ depending
on whether one takes ${\cal A}=0$ or ${\cal B}=0$ in (\ref{fsol}), since the zeros of $J_2$ and
$Y_2$ do not coincide. For an increasing ($a > 0$)/decreasing ($a < 0$) warp factor, the
ghost-like field $\sigma (t,r)$ and the metric function $u(t,r)$ vanish at a finite/infinite
number of points $r_m$ along the extra dimension in the bulk (nodes), forming 4D space-time
`islands' at these nodes, where the matter particles are assumed to be bound.

The induced 4D Einstein space-time is locally $AdS_4$ (in a vicinity of the positions of the
island universes $r_m$ along the extra dimension) if:
\begin{equation}
^{4}R_{ab} = \alpha(r_m) q_{ab}~, ~~~~~ a,b=0,1,2,3
\end{equation}
where $^{4}\!R_{ab}$ and $q_{ab}$ are the induced 4D Ricci and metric tensors, respectively, and
$\alpha(r_m) > 0$ is a constant depending on the position of the island $r_m$.

By computing the 4D induced metric on the brane islands we can write down the components of the 4D
induced Ricci tensor as functions of the induced metric tensor and its first two derivatives, i.e.
as functions of $u(t,r)$ and its first two partial derivatives with respect to time and the extra
coordinate. By looking for a proportionality between the induced metric and Ricci tensors on the
island universes, i.e. when $u(t,r_m) = 0$ (however, the derivatives of $u$ at $r_m$ do not
vanish), we see that the $AdS_4$ effective character of the 4D island universes can be reached
only in two distinct cases:

i) by restricting the amplitude of the anisotropic oscillations to be very small, more precisely
when $AC << 1$, for the islands which are located at finite $r_m$.

ii) asymptotically along the fifth dimension ($r_m \longrightarrow \infty$).

Thus, even when at the $r_m$ points where the island universes are located we have $u(t,r_m)=0$
and the metric (\ref{metric1}) reduces to the standard 5D thin braneworld metric of \cite{brane},
whereas the setup (\ref{action}) simplifies to an action describing 5D gravity with a cosmological
constant, plus a delta-function brane as in \cite{brane}, leading to the known Randall-Sundrum
braneworld which is an $AdS_5$ slice, in general, the metric in the islands (i.e. for each
particular node $r_m$) is not locally $AdS_4$, but just in the above referred two cases.

Returning to the problem of junction conditions (\ref{match}), if we use (\ref{separation}) and
(\ref{fsol}) to solve them, we obtain:
\begin{eqnarray} \label{tensions}
\lambda_0&=&-\frac{3a}{4\pi G_5}~, \nonumber \\
\lambda_1&=& \lambda_2 ~ = \frac{AC\omega\,\epsilon(a)\sin(\omega t)}
{16 \pi G_5}\left( J_3-J_1\right)|_{\omega /|a|}-
\frac{3a}{4 \pi G_5}~, \nonumber \\
\lambda_3&=& -\frac{AC\omega\,\epsilon(a)\sin(\omega t)}{8 \pi G_5}
\left( J_3-J_1\right)|_{\omega /|a|}-
\frac{3a}{4 \pi G_5}~, \label{match1} \\
\,[u']&=& \frac{16 \pi G_5}{3}(\lambda_1-\lambda_3)=AC\omega \,\epsilon(a)\sin(\omega t)\left(
J_3-J_1\right)|_{\omega /|a|}~. \nonumber
\end{eqnarray}
From these relations it can be seen that $\lambda_0 = \rho$ is constant, and it can be interpreted
as the energy density of the 'effective fluid' as mentioned above, while the stresses oscillate in
the $x$, $y$ and $z$ directions with the same frequency as the function $u$ does, as it was
expected since it is precisely along these directions that the metric coefficients do oscillate in
the bulk. On the other hand, the spatial components of the energy--momentum tensor on the brane
can be consistently interpreted following a systematic covariant analysis developed from the
viewpoint of a brane-bound observer \cite{Maartens,KG}.

Let us decompose the tension components (\ref{tensions}) as follows:
\begin{equation}
\lambda_i =-p+ \pi_i~, ~~~~~ (i=1,2,3)
\end{equation}
where the quantity $p$ represents a constant isotropic pressure and the $\pi_i$ are the components
of the anisotropic stress that oscillate along the spatial direction $x^i$:
\begin{eqnarray} \label{anisotropic_t}
p&=&\frac{3a}{4 \pi G_5}~, \nonumber \\
\pi_1&=& \pi_2 ~ = \frac{AC\omega\,\epsilon(a)\sin(\omega t)}{16 \pi G_5}
\left( J_3-J_1\right)|_{\omega /|a|}~,  \\
\pi_3&=& -2\pi_1 = -\frac{AC\omega\,\epsilon(a)\sin(\omega t)}{8 \pi G_5}
\left( J_3-J_1\right)|_{\omega /|a|}~. \label{match1} \nonumber
\end{eqnarray}
By taking into account that in the solution for the junctions conditions (\ref{tensions}) the
parameter $\omega /|a|$ is a zero of $J_2$, but not of $J_1$ and $J_3$, in general the anisotropic
stresses are different from zero. Furthermore, the stresses along the directions $x$ and $y$ are
equal, whereas the stress in the $z-$direction $\pi_3$ is twice in amplitude and possesses
opposite phase with respect to $\pi_1$. This is a direct consequence of the symmetry of the metric
under the exchange $x\longleftrightarrow y,$ on the one hand, and the differences in amplitude and
phase in the argument of the exponential metric coefficients, on the other hand.

Let us compare the isotropic and the anisotropic parts of (\ref{tensormixto}). In order to do that
we shall consider the following physically interesting limits:

\vskip 3mm

\noindent \textbf{a)} The first case is $\omega / |a| \approx 1$, and depending on the amplitudes
$A$ and $C$ in the expressions for the components $\lambda_i$, the dominant term can be the
oscillatory or the constant one ($3a /4 \pi G_5$), or both terms can be of the same order.

\textbf{1a)} If $\omega/|a| \approx 1$ and $AC \ll 1$ the constant term dominates and $\lambda_i
\approx \lambda_0 = - 3a/4\pi G_5$. Then the oscillatory terms in $\lambda_i$ can be neglected
compared to the contribution given by the constant terms and the thin brane becomes "isotropic",
recovering the standard thin brane case \cite{brane}, where the state equation is $p=-\rho$. Thus,
the energy-momentum tensor on the brane only has a relevant contribution coming from the tension
of the brane.

\textbf{2a)} If $\omega/|a| \approx 1$ and $AC\gg 1$ the oscillatory terms are dominant in
$\lambda_i$ and $\mbox{max}(\lambda_i) \gg \lambda_0$, implying that the energy-momentum tensor of
the thin brane describes a kind of exotic matter given by the following expression:
\begin{eqnarray}
\tau^{\mu}_{\nu} \approx \delta{(r)}~\mbox{\rm diag}
[\lambda_0,\tilde{\lambda}_1+\lambda_0,\tilde{\lambda}_2+\lambda_0, \tilde{\lambda}_3+\lambda_0,0]~,
~~~~~ \tilde{\lambda}_1=\tilde{\lambda}_2~,\nonumber
\end{eqnarray}
where
\begin{eqnarray}
\tilde{\lambda}_{1}&=& \frac{AC\omega\,\epsilon(a)\sin(\omega t)}{16 \pi G_5}
\left( J_3-J_1\right)|_{\omega /|a|},\nonumber \\
\tilde{\lambda}_{3}&=&- \frac{AC\omega\,\epsilon(a)\sin(\omega t)}{8 \pi G_5}
\left( J_3-J_1\right)|_{\omega /|a|}.
\end{eqnarray}
In this case, the matter is highly anisotropic because the amplitude of the stationary wave is
large.

\textbf{3a)} If $\omega/|a| \approx 1$ and $AC \approx 1$, the oscillatory and constant terms are
of the same order and $\lambda_i \sim \lambda_0 \sim a$. Due to this fact, in this case there is
no way to highlight an interesting physical situation, in contrast to cases \textbf{1a)} and
\textbf{2a)}.

\vskip 3mm

\noindent \textbf{b)} The second interesting limit takes place when the stationary wave that
propagates in the bulk oscillates in the high frequency regime ($\omega /|a| \gg 1$); in this case
we have again three situations:

\textbf{1b)} If $\omega /|a| \gg 1$ and $AC \ll 1$ with $ACX_{n}\rightarrow C_1,$ where $C_1$ is a
finite constant, we have the same result as in the case \textbf{3a)}. Thus, the oscillatory and
constant terms are of the same order.

\textbf{2b)} If $\omega /|a| \gg 1$ and $AC \approx 1$ we have the same result as in the case
\textbf{2a)}: the matter fluid is highly anisotropic. But in contrast to \textbf{2a)} in this case
it is because the frequency of the stationary wave is large, therefore, it contributes to the
growth of the anisotropic part of the energy-momentum tensor.

\textbf{3b)} If $\omega /|a| \gg 1$ and $AC\gg 1$ we have the same result as in \textbf{2a)}: an
oscillatory energy-momentum tensor of the thin brane which describes a kind of exotic matter. In
this case this fact takes place because the frequency and amplitude of the stationary wave are
large.

Note that since $\omega /|a|$ is a zero of $J_2$ the quantity $\omega/|a|>1$ and the third case
$\omega /|a| \ll 1$ is excluded from  our analysis.

In summary, the two most relevant limits are:

I) when the matter on the brane hardly oscillates due to the small amplitude of the anisotropies
compared to the brane's tension, leading to a quasi-isotropic brane. This limit mimics the
Randall-Sundrum model.

II) when the matter on the brane is highly anisotropic and a stationary wave propagates through
the bulk. In this case we are far from the Randall-Sundrum limit.

It would be interesting to study a braneworld isotropization mechanism for our brane universe
similar to those proposed in \cite{anisotbwinfl,bwisotropization}. In the latter model, the brane
anisotropic energy leaks into the bulk as it evolves, a phenomenon that can also be interpreted
from a completely 4D point of view in the framework of the AdS/CFT correspondence as particle
production in the CFT, where energy is drawn from the anisotropy to fuel the process, leading to
an isotropic brane over time. In principle, this study could be carried out for our anisotropic
braneworld model since we have completely solved the 5D Einstein equations, a necessary
requirement for the computation of the projection of the Weyl tensor into the brane.


\section{Localization of transverse scalar and metric fields}

In the original paper \cite{Gog1} it was shown that near the nodes $r_m$, where $u(r,t)\approx 0,$
the metric (\ref{metric1}) adopts the usual thin brane form of \cite{brane} and one recovers
localized 4D gravity on the island universes in the usual way through a massless mode solution for
the equation that governs the dynamics of transverse traceless metric fluctuations.

Here we shall consider the localization of small perturbations of a real massless scalar field
defined by the action:
\begin{equation} \label{Sphi}
S_{\varphi}= -\frac{1}{2} \int \sqrt{g} dx^{4}dr \ g^{\mu \nu}\partial_{\mu}\varphi
\partial_{\nu}\varphi~.
\label{actionphim0}
\end{equation}
We consider a massless scalar field since it is important to get an oscillating solution in our
model, otherwise there will be no standing wave since the solution for the scalar field $\varphi$
will be dissipative (as in the case of including a mass term, for instance). It is necessary to
mention that unlike \cite{Gog1} we study the localization of scalar field in the system formed by
the thin brane located in $r=0$ and all the island universes. This will be seen below from the
fact that the norm of fluctuations is defined on the whole domain of the extra dimension $r$,
thus, if it is finite for the entire domain, it will also be finite for single islands or a set of
island-universes.

The scalar field equation corresponding to the action (\ref{Sphi}) is:
\begin{equation} \label{fieq}
\ddot\varphi - e^{-u}(\varphi_{xx}+\varphi_{yy}) - e^{2u}\varphi_{zz} - e^{2 a |r|}
\left[4 a \epsilon(r)\varphi' +
\varphi''\right] = 0~,
\end{equation}
where the subscripts $xx$, $yy$ and $zz$ denote second differentiation with respect to $x$, $y$
and $z$, respectively. Let us propose the following {\it ansatz} for the scalar field,
\begin{equation}
\varphi=L(r,t)P_1(x)P_2(y)P_3(z)~,
\end{equation}
which transforms the previous equation into,
\begin{equation} \label{fieq2}
\frac{\ddot L}{L} - e^{-u}\left( \frac{1}{P_1}\frac{d^2P_1}{dx^2}+
\frac{1}{P_2}\frac{d^2P_2}{dy^2}\right) -
\frac{e^{2u}}{P_3}\frac{d^2P_3}{dz^2} - e^{2 a |r|}
\frac{\left[4 a \epsilon(r)L' + L''\right]}{L} = 0 ~.
\end{equation}
Let us also set
\begin{equation} \label{eqsP}
\frac{d^2P_i}{dx^2_i} = -k_i^2P_i~, ~~~~~ (i=1,2,3.)
\end{equation}
The parameters $k_i$ can be physically regarded as the momenta along the $i$ direction of the
brane without considering the anisotropic character of the background metric.

By taking into account (\ref{eqsP}), we get:
\begin{equation} \label{fieq3}
\ddot L - e^{2 a |r|}\left[ 4 a \epsilon(r)L' + L''\right]= -\left[ (k^2_1+k^2_2)e^{-u}
+k^2_3e^{2u}\right] L ~.
\end{equation}
In the limit $u \rightarrow 0$ this equation has the well known solution:
\begin{equation}
L \sim e^{ik_0t}~, ~~~~~ k_0^2 - k_i^2 = 0~,
\end{equation}
corresponding to the 4D massless scalar mode, which can be localized on the thin brane located at
$r=0$ by the warp factor $e^{2a|r|}$ if $a<0$ (see, for example, \cite{local}).

In this paper we want to consider a simple case of transverse fluctuations to the island
universes, i.e. when $k_i\approx 0$. In this case the equation (\ref{fieq3}) takes the form:
\begin{equation}
\label{eqL} \ddot L - e^{2 a |r|}\left[4 a \epsilon(r)L' + L''\right] = 0 ~.
\end{equation}
By proposing a similar to (\ref{separation}) oscillatory {\it ansatz}:
\begin{equation}\label{L}
L(r,t)=K\sin(\Omega t)g(r)~,
\end{equation}
where $\Omega$ and  $K$ are some constants, we obtain,
\begin{equation}
\label{g} g'' + 4 a \epsilon(r) g' + \Omega^2 e^{-2 a |r|}g = 0~,
\end{equation}
which is the same equation as (\ref{f}) for the metric function $f(r)$, and thus, possesses the
solution:
\begin{equation} \label{gsol}
g(r) = e^{-2a|r|} \left[D~J_2\left( \frac{\Omega}{|a|} e^{-a|r|}
\right) + E~Y_2\left( \frac{\Omega}{|a|} e^{-a|r|} \right)\right],
\end{equation}
where $D,E$ are integration constants. In contrast with the frequency of the ghost-like field,
$\omega$, the parameter $\Omega$ in (\ref{gsol}) is not quantized since in general $\Omega \neq
\omega$.

In order to analyze the proper way of normalization of $g(r)$, we rewrite (\ref{g}) as:
\begin{equation} \label{eqf}
-\left(e^{4a|r|}g'\right)'=\Omega^2 e^{2 a |r|} g ~.
\end{equation}

Now we recall the Sturm--Liouville method associating to the equation,
\begin{equation} \label{STp}
-\left[p(r)y'\right]'+q(r)y=\lambda s(r)y~,
\end{equation}
of a variable $y(r)$, the norm,
\begin{equation} \label{norm}
||y||^2_s=\int_b^c |y(r)|^2s(r)dr~,
\end{equation}
where $\lambda$ is an eigenvalue parameter, while $b$ and $c$ are arbitrary real constants. By
comparing (\ref{STp}) with (\ref{eqf}) we see that $y(r)=g(r)$, $q(r)=0$, $s(r)=e^{2 a |r|}$,
$\lambda=\Omega^2$, $p(r)=e^{4 a |r|}$ and the correct norm for (\ref{g}) is given by,
\begin{equation} \label{normg}
\int_b^c |g(r)|^2e^{2a|r|}dr~,
\end{equation}
where $b=0$ and $c=\infty$. Thus $g(r)$ belongs to the Hilbert space $H_s$ consisting of all such
functions for which (\ref{normg}) is finite \cite{stakgold}.

Another way of looking at this problem consists of considering the action for the scalar field
(\ref{actionphim0}) and taking into account (\ref{metricA}) and the expression for $\sqrt{g}$, in
order to show that its nontrivial part (the kinetic term) is,
\begin{equation}
S_{\varphi}\sim \int dx^{4}dr e^{2a|r|} \dot{L}^{2} + \cdots~,
\end{equation}
where the dots denote the 5D contribution of the scalar field. If we further make use of the
relation (\ref{L}), then the normalization condition along the extra coordinate reads:
\begin{equation} \label{localization}
S_{\varphi} \sim \int dr |g(r)|^{2}e^{2a|r|}~,
\end{equation}
which precisely coincides with relation (\ref{normg}). Thus, if we want to have a localized
transverse 5D scalar field on the brane (and due to relation (\ref{sigma=u}), the gravitational
waves $u$ as well) the integral over $r$ in (\ref{localization}) must be finite.

It should be pointed out here that the finiteness of the above obtained norm (\ref{normg}) is
exactly the same for the phantom-like scalar field ($\phi$, or $\sigma$), which indeed, vanishes
at the island branes. Therefore, the latter field will be localized as well if the expression
(\ref{localization}) constitutes a finite quantity.

By considering the change of variable $v=\Omega e^{-a|r|}/|a|$ for the case $a>0$ we have the
following integrating limits $v_1 = \Omega /|a|$ and $v_2 = 0$. Alternatively, for $a<0,$ we get
$v_1 =\Omega/|a|$, while $v_2 =\infty$.

In the language of this new coordinate, the integral
(\ref{localization}) adopts the form:
\begin{eqnarray} \label{besseljy}
S_{\varphi} \sim -\int _{v_1}^{v_2} v \left[|D|^2~ |J_2(v)|^2+
D\overline{E}~J_2(v)\overline{Y_2(v)} \right. +  \nonumber \\
 + \left. \overline{D}E~\overline{J_2(v)}Y_2(v) + |E|^2~
|Y_2(v)|^2\right]dv~,
\end{eqnarray}
where $\overline{X}$ denotes the complex conjugate of $X$. It is easy to get the behavior of the
integral (\ref{besseljy}) by direct computation. It turns out that only in the case $E=0$ and
$a>0$ this integral is finite. Therefore the massless scalar field $\varphi$ is localized on the
brane at $r=0$, as well as in all the island branes of the model since in this latter case one
should just change the integration limits $v_1$ and $v_2$, getting a finite value for each
island-universe or for a set of them.

It should be mentioned that if one compares our result to the result obtained in
\cite{Gog1,local}, it seems that they are in an apparent contradiction, because the scalar field
is localized for $a>0$ in our case, while in their papers this field is localized for $a < 0$.
However, the point here is that we are investigating the transverse modes and not the modes along
the brane. Also in our work $\Omega\neq 0$ in the \emph{ansatz} (\ref{L}), otherwise $L\equiv 0$
and the scalar field $\varphi$ would be trivial. A way of avoiding this trouble consists in
extending the \emph{ansatz} (\ref{L}) to contain a nonzero constant phase $\alpha \neq 2\pi n$
with $n \in Z$ in the argument of the sinus function, i.e.
\begin{equation}
L = K\sin\left(\Omega t+\alpha\right)g(r)~.
\end{equation}
In this case the $\Omega=0$ value corresponds to the lowest frequency of oscillation of the
nontrivial mode of the scalar field $\varphi$, which indeed adopts a constant value along the
fifth dimension (as well as $g$). Then this field is localized for $a < 0$ like in
\cite{Gog1,local} according to the normalization condition (\ref{normg}).

Now let us roughly analyze the gravitational sector from this localizing point of view. The Ricci
scalar for the metric (\ref{metric1}) reads:
\begin{equation} \label{ricci5}
R_{5}= -\frac{3}{2} e^{-2a|r|}\dot{u}^{2} +\frac{3}{2}u'^{2}+20a^{2}+16a\delta(r)~.
\end{equation}
Thus, the gravitational 5D action is:
\begin{equation}
S_{g}=\int \sqrt{g} dx^{5} R_{5} \sim - \frac{3}{2} \int dx^{5} e^{2a|r|}\dot{u}^{2} + \cdots~,
\end{equation}
where dots denote 5D contributions. Due to the relation (\ref{sigma=u}) between the phantom-like
scalar field $\sigma$ and the gravitational field $u$, it turns out (as expected) that
localization properties of these fields are similar. Thus the gravitational field $u$ is also
localized if we choose $E=0$ and $a>0 $ in the solution (\ref{gsol}) under the {\it ansatz}
(\ref{L}).


\section{Final Remarks}

In this paper we presented a consistent derivation of the 5D anisotropic standing wave braneworld
proposed in \cite{Gog1} by initially assuming a $Z_2$--symmetric factor and introducing a simple
anisotropic energy-momentum tensor with different stresses along different space-time directions
on the $3$-brane. We also derived and explicitly solved the corresponding junction conditions for
the braneworld model and obtained analytical expressions for the (oscillating) stresses of the
$3$-brane along the 4D spatial directions.

By carefully looking at the energy-momentum tensor of the $3$-brane proposed in
(\ref{tensormixto}) we can infer from (\ref{tensions}) that it corresponds to an anisotropic
effective fluid which is a mixture of a vacuum fluid characterized by its tension and an
anisotropic oscillating matter fluid. We can also conclude that the 4D matter corresponding to
this source has an exotic nature because it violates the weak energy condition for increasing warp
factors and satisfies it for decreasing warp factors ($a<0$). It is worth mentioning here that
this condition is also violated in the thin brane models \cite{brane}. Besides, it violates as
well the dominant energy condition since it is an oscillating fluid which emits anisotropic waves
into the bulk with different amplitudes (which change sign over time and eventually disappear) and
phases along the directions $x$, $y$ and $z$. Nevertheless, one can look for an alternative
mechanism that could lead to a more physical picture and eventually heal this drawback related to
the anisotropy of the metric {\it ansatz} (\ref{metric1}), like proposing a more involved
energy-momentum tensor or add some matter fields on the $3$-brane. An interesting situation takes
place when the amplitude of the anisotropies is small with respect to the tension on the brane
(quasi-isotropic limit), since in this case the braneworld is effectively isotropic. Moreover,
within the framework of the model presented here it is possible to study the braneworld
isotropization in which anisotropies dissipate via inflation \cite{anisotbwinfl} or leakage of
thermal graviton radiation into the bulk \cite{bwisotropization}, a relevant phenomenon from the
brane cosmological viewpoint that deserves more attention.

We showed as well the correct way of defining the norm for the transverse scalar field $\varphi$
through the Sturm-Liouville method, a fact that leads to the localization of this field just when
$a>0$ and $E=0$ in the solution (\ref{gsol}). However, it should be pointed out that this
localization approach involves all the island brane universes as well as the thin brane located at
$r=0$. While the approach taken in \cite{Gog1} considers the gravity localization on each of the
island universes and is related to the massless modes of the transverse traceless metric
fluctuations of the system, i.e. to the 4D gravitons that live in each $3$-brane. Our results seem
to agree with recent results obtained in \cite{GK}, where localization of massive test particles
about a thick braneworld with a time dependent extra dimension arises just for increasing warp
factors due to an oscillatory behaviour in time--like geodesics. Further work must be done in
order to get localization mechanisms of matter fields with decreasing warp factors, since these
are useful in solving the hierarchy problem. We also would like to point out that further
investigations towards the localization of three generations of fermion fields in this model, both
at the thin brane located at $r=0$ and at the island universes located at $r_m$, are in progress
and will be published elsewhere.

We finally would like to point out that in order to avoid the instabilities related to the
ghost-like scalar field of this braneworld we can appeal to an alternative approach of
interpreting the phantom--like scalar field $\phi$ as the geometrical scalar field of a 5D Weyl
integrable manifold \cite{Weyl1,Weyl2}, where a scalar appears through the definition of the
covariant derivative of the metric tensor,
\begin{equation} \label{D}
D_{\gamma} g_{\alpha\beta} = g_{\alpha\beta}\partial_\gamma \phi ~.
\end{equation}
This is a generalization of the Riemannian geometry, in which the covariant derivative of the
metric tensor obeys the metricity condition, i.e. it vanishes
\begin{equation} \label{DR}
D_{\gamma} g_{\alpha\beta} =0 ~.
\end{equation}
On the other hand, this relation indicates that the Weylian affine connections are not metric
compatible since they also involve the scalar field in their definition:
\begin{equation}
\label{affconnect} \Gamma_{\mu\nu}^\rho=\{_{\mu\nu}^{\;\rho}\}-\frac{1}{2}\left(
\phi_{,\mu}\delta_\nu^\rho+\phi_{,\nu}\delta_\mu^\rho-g_{\mu\nu}\phi^{,\rho}\right)~,
\end{equation}
where $\{_{\mu\nu}^{\;\rho}\}$ are the Christoffel symbols. As a consequence, in an integrable
Weyl manifold specified by the pair $(g_{\mu\nu},\phi)$, the non-metricity condition (\ref{D})
implies that the length of a vector is altered by parallel transport.

The 5D Weyl action can be written as
\begin{equation}
\label{Waction} S_5^W =\int_{M_5^W}\frac{d^5x\sqrt{|g|}}{16\pi
G_5}e^{-\frac{3}{2}\phi}\left[R+3\tilde{\xi}\left(\nabla\phi\right)^2-2U(\phi)\right]~,
\end{equation}
where $M_5^W$ is a Weylian integrable manifold, $\tilde{\xi}$ is an arbitrary coupling parameter
and $U(\phi)$ is a self-interaction potential for the scalar field $\phi$. From the formalism
itself it becomes clear that this action is of pure geometrical nature since the scalar field that
couples non-minimally to gravity is precisely the scalar $\phi$ that enters the definition of the
affine connections of the Weyl manifold (\ref{affconnect}) and the non-metricity condition
(\ref{D}) and, thus, cannot be neglected at all in our setup.

By performing the conformal transformation,
\begin{equation}
\label{conftransf} \widehat{g}_{\mu\nu}=e^{-\phi}g_{\mu\nu}~,
\end{equation}
we map the Weylian action (\ref{Waction}) into the Einstein frame:
\begin{equation}
\label{confaction} S_5^R=\int_{M_5^R}\frac{d^5x\sqrt{|\widehat
g|}}{16\pi G_5}\left[\widehat R+3{\xi}\left(\widehat\nabla\phi\right)^2-2\Lambda_5\right]~,
\end{equation}
where all hatted magnitudes and operators are defined in the Riemann manifold,
$\xi=\tilde{\xi}-1$, $\ \widehat U(\phi)=e^{-\phi} U(\phi)=\Lambda_5$ and we have set
$U(\phi)=\Lambda_5e^{\phi}$ in (\ref{Waction}) in order to get a 5D cosmological constant in
(\ref{confaction}) as in (\ref{action}). In this frame we have a theory which describes 5D gravity
minimally coupled to a scalar field plus a cosmological constant, the affine connections become
the Christoffel symbols, the metricity condition is recovered and Weyl's scalar field in
(\ref{confaction}) imitates a massless scalar field (either an ordinary scalar or a ghost-like
scalar depending on the sign of $\xi$ \cite{Weyl1,Weyl2}).

In the alternative approach mentioned above we can start with the action (\ref{Waction}) which
does correspond to a conventional scalar field in the Einstein frame's action (\ref{confaction})
under the conformal transformation (\ref{conftransf}). Then we can compute the corresponding field
equations under the metric (\ref{metric1}) (which will be different from (\ref{field-eqnsDELTA})
since the affine connections involve the scalar field) and solve them in the Weyl frame. We
further can set $\phi_{,\mu}=0$ on the brane in the solution of the field equations (which also
will be different from the solution in the Riemann manifold). Thus, in this way we can recover an
effective 4D universe in the Einstein frame at the brane (and possibly at the island universes if
our solution possess the same structure as the braneworld obtained in \cite{Gog1}) completely free
of ghost instabilities since the Weyl scalar corresponds to a conventional real scalar field. This
scenario still has to be investigated in full detail, a direction which is currently under
research.


\section{Acknowledgements}

We thank P. Midodashvili and U. Nucamendi for reading the manuscript and giving us their comments.
This research was supported by grants CIC-4.16 and CONACYT 60060-J. DMM acknowledges a PhD grant
from CONACYT. AHA thanks SNI for support. MG was supported by the grant of Rustaveli National
Science Foundation $ST~09.798.4-100$ and by the research project CONACYT 60060--J as well.


\end{document}